# Super pulses of orbital angular momentum in fractional-order spiroid vortex-beams

A. V. Volyar*, Yu. A. Egorov

*V.I. Vernadsky Crimean Federal University, Vernadsky Prospekt, 4, Simferopol, 295007, Russia*
*Corresponding author: volyar@crimea.edu*

We consider optical properties of singular beams with spiral-like intensity and phase distributions that were called the spiroid beams. Their orbital angular momentum as a function of a fractional-order topological charge has a chain of supper-pulses (bursts and dips). The form of the supper-pulses can be controlled by the spiral parameters, Such a phenomenon can be used in optical switches and triggers for optical devices and communication systems.

The first investigations of vortex-beams with fractional-order topological charges presented by Soskin et. al. in [1] made a lot of physicists to peer more attentively into physical features of such remarkable optical constructions. Later, a simple physical treatment of optical processes in such a complex beam structure was considered by Berry [2] that stimulated a torrent of publications on the problem (see, e.g. [3-16] and references therein). Berry showed that a fractional-order vortex-beam can be treated as an infinite number of singular beams with negative and positive integer-order topological charges but different energy distribution. Then it became clear why a fractional-order vortex beam can bear a fractional orbital angular momentum (OAM) [3,4]. In connection with it a special attention was focused on the vortex spectral content for controlling values of OAM [5-7] and for shaping a wished form of focal traps for manipulations of microparticles or for light communication systems [8,9].

The propagation properties of vector half-order vortex-beams in free space and crystals were presented in the papers [10,11]. Moreover it was revealed [12] that a circular fiber array can serve as a medium for propagation of half-integer order super modes without their decay into integer-order beans (as it was predicted in [2]) while all other integer order ones are a superposition of half-integer-order vortex-modes.

The OAM distribution in fractional-order vortex-beams depends essentially on the beam types. The theoretical consideration of the problem was presented in the papers [13,14] for Bessel-Gaussia and Laguerre-Gaussian - type beams. Authors remarked that the OAM follows the vortex topological charge with small oscillations for low-order vortices. This was experimentally confirmed in the paper [19]. However the detailed analysis showed [15] that there are OAM pulses with large amplitudes for higher-order vortices near integer-order values.

Note, that variations of vortex content in fractional-order vortex- beams have an influence upon a form of the bursts and dips in the OAM contour and to act as a switch in different optical devices. A similar assumption has been suggested by the authors of the paper [16] that considered a helico-conical optical beam – one of representatives of fractional-order light fields. They revealed experimentally that a superposition of beams with helical and conical wavefronts obtained on the base of a higher-order Bessel beam (diffracted by a spatial phase modulator) forms a complex field with the one-branch spiral intensity distribution at the beam cross-section. Variations of the beam parameters result in transformations of the spiral form.

In the letter we generalized such a representation employing the fractional-order hypergeometric-Gaussian (Hy-G) beams with variations of their parameters that enable us to reconstruct smoothly the shape of bursts and dips in the OAM contour.

Thus, the aim of our paper is to shape the fractional-order Hy-G vortex-beam and to modify its spectral vortex content so that to control the OAM distribution among the integer-order vortices.

We start with the angular spectrum of plane monochromatic waves written in the form at the initial plane $z=0$

$$U = \frac{k_\perp^{|p|}}{2}\left[e^{i\psi_1} + e^{-i\psi_2}\right]e^{-\frac{k_\perp^2 w_0^2}{4}} = \frac{1}{2}(U_1 + U_2), \quad (1)$$

where

$$\psi_1 = a_1 k_\perp^2 - p\phi, \quad \psi_2 = a_2 k_\perp^2 - p\phi. \quad (2)$$

$k_\perp$ is a transverse wave number, $p$ is the fractional-order topological charge, $\phi$ stands for the azimuth angle in $\mathbf{k}$–space. The phases $\psi_{1,2}$ describe the spiral-like (spiroid) phase distribution in the plane wave spectrum, while parameters $a_{1,2}$ (arbitrary real numbers) define torsion of the spirals, $w_0$ is a beam waist radius.

**1**. The scalar field distribution in the polar coordinates $(r,\varphi,z)$ is found by the Fourier transform

$$\Psi_1(r,\varphi,z,a_1) = \frac{1}{2\pi}\int_0^\infty k_\perp dk_\perp \int_0^{2\pi} U_1 e^{ik_\perp r\cos(\phi-\varphi)+iz\sqrt{k^2-k_\perp^2}} d\phi .\quad(3)$$

In the paraxial approximation $\sqrt{k^2-k_\perp^2} \approx k - k_\perp^2/(2k)$, where $k$ is a wavenumber, we find

$$\Psi_1 = -\frac{i}{\pi}\sin p\pi\, e^{-ip\pi} \sum_{m=-\infty}^{\infty} i^m \frac{e^{im\varphi - \frac{\rho^2}{\sigma_1}+ikz}}{p+m} \frac{\rho^{|m|}}{2^{|m|}\sigma_1^{\frac{|p|+|m|+2}{2}}} \times \qquad (4)$$

$$\times \frac{\Gamma\big((p+2+|m|)/2\big)}{|m|!}\,{}_1F_1\left(-\alpha;\beta;\frac{\rho^2}{\sigma_1}\right),$$

where $\alpha = \frac{|m-p|}{2}$, $\beta = |m|+1$, $\rho = r/w_0$, $\Gamma(x)$ is a Gamma function, ${}_1F_1(a;b;x)$ is a hypergeometric function, $\sigma_1 = 1 + i(z-2ka_1)/z_0$, $z_0 = kw_0^2/2$.

Each tern in Eq.(4) needs normalizing because of the contribution of the Hy-G beam to the total energy flow rises quickly as the index $m$ increases. The normalizing factor is $N_{m,a}^2 = (1+\bar{a}_1^2)^{|m|}\Gamma(|m|+1)/2^{|m|}$, $a_1 = \bar{a}_1 \cdot 10^7$, where we took into account that ${}_1F_1(0,\beta,x) = 1$. The normalized intensity distributions as a function of topological charges $p = m$ illustrates **points** in Fig.1 for different values of the spiral torsion $\bar{a}_1$.

The above normalization is valid only if each term in Eq.(4) obeys the paraxial wave equation. Indeed, authors of the paper [17] showed that the Hy-G beams are the solutions to the paraxial wave equation and analyzed different beam types while the authors of the paper [18] produced binary diffractive optical elements for generating these beams,

The intensity and phase distributions shown in Fig.2 demonstrate a complex structure of the fractional order Hy-G beams with different values of the topological charge at the plane $z = 0$. The speckled intensity pattern for the charge $p = 3.5$ transforms into a regular ring structure at the charge $p = 3,9$ forthcoming the integer-order $m = 4$. However, in contrast to the standard vortex beam we observe bending of the nodal lines into spiral-like (spiroid) structure. Although the hypergeometric function in Eq.(4) transforms into unit at $p \approx m$ so that radial intensity distribution tends to that of the standard vortex. The phase structure embalms nevertheless its memory about spiral-like phase structure of the angular spectrum. As we will make sure later the above beams have a line of properties unlike the standard vortex- beams. We called them the *spiroid beams*.

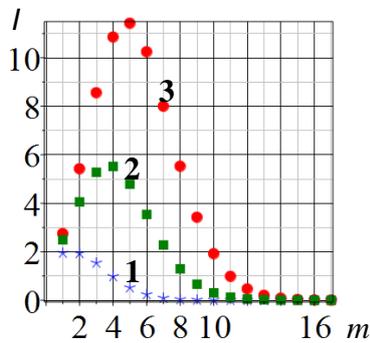

Fig.1 Intensity distribution $I(m)$
$p = m$: **1** $-\bar{a}_1 = 0.5$,
**2** $-\bar{a}_1 = 1$; **3** $-\bar{a}_1 = 2$;
$w_0 = 0.5mm$, $z = 0$

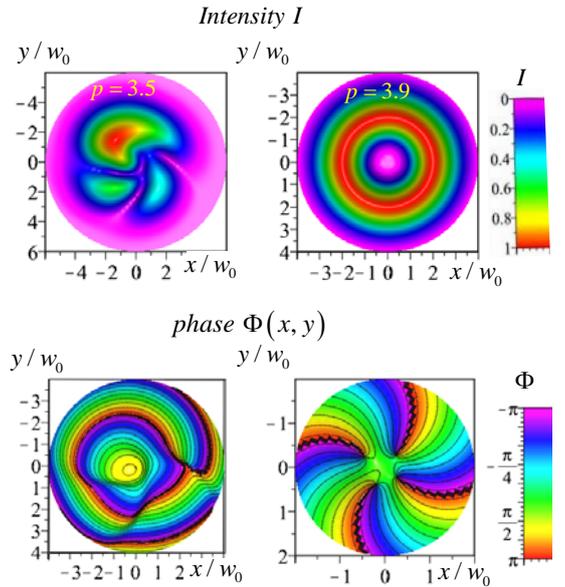

Fig.2 Intensity and phase distributions of the vortex-spiroid beam.
($w_0 = 1mm$, $\bar{a}_1 = 1.5$, $z = 0$)

The phase evolution $\Phi(x,y,z)$ shown in Fig.3 illustrates strange phase behavior. Really, at the position $z = 0$ we observe the right-hand spiral twist. At the position $z = 8m$ the spiral changes the handedness while at the position $z = 4$, the spiral vanishes. Instead we observe three helicoids formed

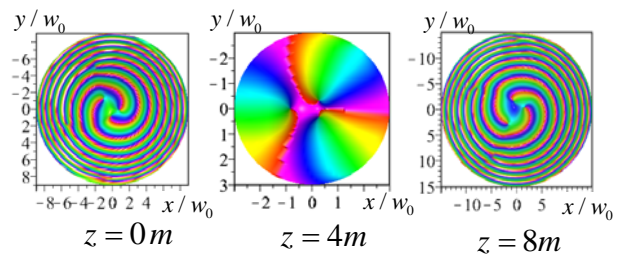

$z = 0m$  $z = 4m$  $z = 8m$

Fig.3 Phase evolution $\Phi(x,y,z)$ along the beam length: $p = 2.9$; $\bar{a}_1 = 2$

by three off-center vortices. The situation clears up if one focuses attention on the parameter $\sigma_1$ (see Eq.(4)) where the spiral torsion $a_1$ provokes the beam waist displacement at the distance $z = 2ka_1$.

**2.** Let us consider the torsion influence $a_1$ on the spiroid beam OAM $L_z$. A simple calculation shows that the specific OAM $l_z = L_z / I$ is equal to

$$l_z = \frac{\sum_{m=-\infty}^{\infty} \frac{m}{(p+m)^2 (1+a_1^2)^{|m|}}}{\sum_{m=-\infty}^{\infty} \frac{1}{(p+m)^2 (1+a_1^2)^{|m|}}} \quad . \quad (5)$$

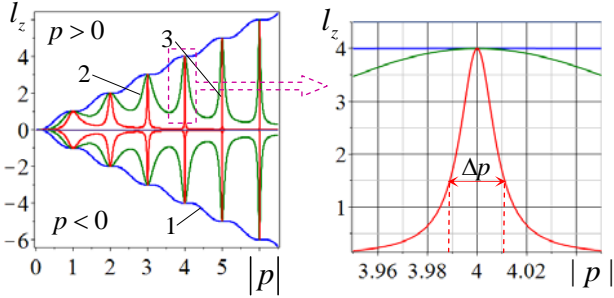

Fig.4 OAM $l_z(P)$ of the vortex- spiroid beam: **1**- $\bar{a}_1 = 0.5$, **2**- $\bar{a}_1 = 2.5$, **3**- $\bar{a}_1 = 5$; $w_0 = 1mm$

Pulsations of OAM shown in Fig.4 have been also observed in the paper [15] for the fractional-order Bessel-Gaussian beams. When the value of the fractional topological charge $p$ changes, there appears a chain of the OAM pulses with their maxima at the integer order values $p = m$. But the pulse shape did not depend on the beam parameters. In our case we can control the shape of the pulse changing the torsion parameter $a_1$. At the small values of the spiral torsion $a_1$, the pulse amplitude is small too (curve-1 in Fig.4). However, a little raise of the $a_1$- parameter induces a chain of super-pulses – "bursts" of OAM with a great amplitude and a very narrow pulse width $\Delta p$ (curve-3 in Fig.4).

**3.** Now we consider a total picture of the beam shaping controlling the OAM supper-pulses hidden in the expression (1). The beam field $\Psi_2(r, \varphi, z, a_2)$ can be found by the transformation of eqs (1) and (4) as $p \to -p, a_1 \to -a_2$ so that we obtain $\Psi(a_1, a_2) = \Psi(a_1) + \Psi(a_2)$.

Intensity and phase distributions in these beams have essential external differences in comparison with the above spiroid beams in Fig. 2. Phase distributions in Fig.5 has spiral-like form inserted into off-center curves. When difference $\Delta a = 0$ the intensity distribution takes the form of a spiral in Fig.5c where six spiral branches are divided by edge dislocations.

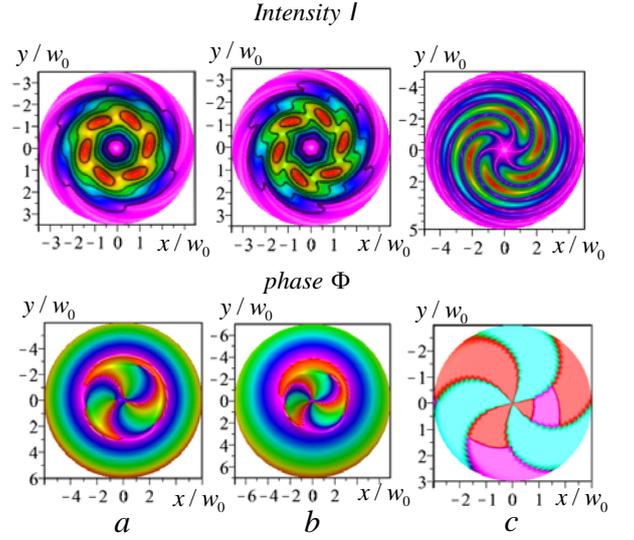

Fig.5 Intensity and phase distributions in the gear-like beams $(w_0 = 0.5mm, \ p = 2.9)$:
$a - \bar{a}_1 = 1, \ \bar{a}_2 = 3$; $b - \bar{a}_1 = 1, \bar{a}_2 = 1.5$;
$c - \bar{a}_1 = \bar{a}_2 = 1$

However, the internal difference between two types of the spiroid beams is concealed in the OAM. The specific OAM $l_z$ of the gear-like beams is

$$l_z = \frac{\sum_{m=-\infty}^{\infty} \left\{ \frac{m}{(p+m)^2 (1+a_1^2)^{|m|}} + \frac{m}{(p-m)^2 (1+a_2^2)^{|m|}} \right\}}{\sum_{m=-\infty}^{\infty} \left\{ \frac{1}{(p+m)^2 (1+a_1^2)^{|m|}} + \frac{1}{(p-m)^2 (1+a_2^2)^{|m|}} \right\}} \quad . \quad (6)$$

The OAM curves shown in Fig.6 illustrate sharp dips near integer numbers $p \approx m$. The burst depth is proportional to the index $m$ and depends on the $a_{1,2}$– parameters but the line width is essentially narrower than that in Fig.4. Between the neighboring bursts, the OAM value is almost zero. The dip width is controlled by the $a_{1,2}$– parameters so that the burst can take the form of the Dirac delta function near arbitrary $p \approx m$ including $m = \pm 1$. Note that the substitution $a_1 \rightleftarrows a_2$ results in converting the beam OAM $l_z \to -l_z$. Such a phenomenon can be used in optical switches and triggers.

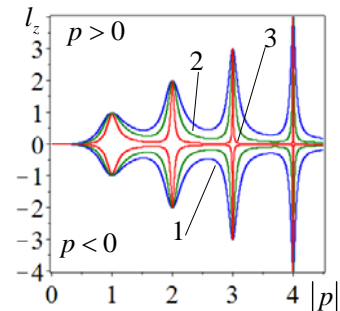

Fig.6 OAM $l_z(p)$ of spiroid beams :
**1** – $\bar{a}_1 = 1, \bar{a}_2 = 7$ ; **2** – $\bar{a}_1 = 2, \bar{a}_2 = 8$ ;
**3** – $\bar{a}_1 = 3, \bar{a}_2 = 10$

1. Basisty, M. Soskin, M. Vasnetsov, Opt. Commun. **119**, 604-612 (1995).
2. M. V. Berry, J. Optics. A **6**, 259-268 (2004).
3. J. Leach, E. Yao, and M. J. Padgett, New J. Phys. **6**, 71 (2004).
4. J. Götte, S. Franke-Arnold, R. Zambrini, and S. M. Barnett, J. Mod. Opt. **54**, 1723–1738 (2007)
5. J. B. Götte, et al., Optics Express **16**, 993-1006 (2008).
6. J. A. Rodrigo, et al., Opt. Express **19**, 6, 5232 (2011).
7. L. Torner, J. Torres, S. Carrasco, Opt. Express 13, 873 (2005).
8. Yi-Dong Liu, C. Gao1,, Xiaoqing Qi, H. Weber, Opt. Express **16**, 7091 (2008).
9. S. Li, J. Wang, Opt. Express **23**, 18736 (2015).
10. T. Fadeyeva, et al., Opt. Lett. **37**, 1397 (2012).
11. T. Fadeyeva, et al., J. Opt. **14**, 044020 (2012).
12. C. Alexeyev, et al., Opt. Lett. **42**, 783 (2017).
13. C. Gutiérrez-Vega, C. López-Mariscal, J. Opt. A, Pure Appl. Opt. **10**, 015009 (2008).
14. Martinez-Castellanos, J. C. Gutiérrez-Vega. J. Opt. Soc. Am. A. **30**, 2395-2400 (2013).
15. T. A. Fadeyeva, et al., J. Opt. Soc. Am. B **31**, 798 (2014).
16. C. A. Alonzo, P. J. Rodrigo, J. Glückstad, Opt. Express **13**, 1749 (2005)
17. E. Karimi, et al., Opt. Lett. **32**, 3053 (2007)
18. V. V. Kotlyar, et al., Applied Optics **47**, 6124 (2008).
19. S. N. Alperin, Opt. Lett. **41**, 5019 (2016).